\documentclass[twocolumn,prd,showpacs,preprintnumbers,amsmath,amssymb]{revtex4}
\usepackage{amsmath,mathrsfs,amsbsy,color,graphicx,bm,amsthm,amsfonts}
\usepackage{units}
\usepackage{bbm}
\usepackage{times}
\usepackage{dcolumn}
\usepackage{mathrsfs}
\usepackage{amsmath,amssymb,epsfig}

\begin{document}

\title{Gaussian quantum steering and its asymmetry in curved spacetime}
\author{Jieci Wang$^{1,2}$\footnote{Email: jcwang@hunnu.edu.cn}, Haixin Cao$^{1}$, Jiliang Jing$^{1}$\footnote{Email: jljing@hunnu.edu.cn}, and Heng Fan$^{2}$\footnote{ Email: hfan@iphy.ac.cn}}
\affiliation{$^1$ Department of Physics, and Collaborative Innovation Center for Quantum Effects \\
and Applications,
 Hunan Normal University, Changsha, Hunan 410081, China\\
 $^2$ Beijing National Laboratory for Condensed Matter Physics, Institute of Physics,Chinese Academy of Sciences, Beijing 100190, P. R. China
}


\begin{abstract}
We study Gaussian quantum steering and  its asymmetry in the background of a Schwarzschild black hole. We present a Gaussian channel description of quantum state evolution under the influence of the Hawking radiation.
We find that thermal noise introduced by Hawking effect will destroy the steerability between an inertial observer Alice and an accelerated observer Bob who hovers outside the event horizon, while it generates steerability between  Bob and a hypothetical observer anti-Bob inside the event horizon. Unlike entanglement behaviors in curved spacetime,
here the steering from Alice to Bob suffers from a ``sudden death" and the  steering from anti-Bob to Bob experiences a ``sudden  birth" with increasing Hawking temperature. We also find that the Gaussian steering is always asymmetric and the maximum steering asymmetry cannot exceed $\ln 2$, which means
 the state never evolves to an extremal asymmetry state. Furthermore,  we obtain  the parameter settings that maximize steering asymmetry and find that (i)  $s=arccosh(\frac{\cosh^2 r}{1-\sinh^2r})$ is the  critical point of steering asymmetry, and (ii) the attainment of maximal steering  asymmetry indicates the transition between one-way steerability and both-way  steerability  for the two-mode  Gaussian state under the influence of Hawking radiation.
\end{abstract}

\vspace*{0.5cm}
\maketitle
\section{Introduction}

Einstein-Podolsky-Rosen steering \cite{schr,schr2},  an intermediate type of quantum correlation between entanglement and Bell nonlocality,
has recently attracted renewed interest \cite{wiseman,Skrzypczyk,Walborn, Bowles,steering1, steering2, steering3, steering4, steering5}.
Steering is a quantum phenomenon that allows one to manipulate the state of one subsystem
by performing measurements on the other entangled subsystem. After being realized
by Schr\"odinger \cite{schr,schr2}, the concept of quantum steering was studied
by Einstein, Podolksy and Rosen (EPR) in their
well-known 1935 paper \cite{epr}, and was treated as the
core concept of the EPR paradox \cite{eprpar}.
The experimental
 detection of quantum steering, i.e.,
 for the demonstration of the EPR paradox,
 was first proposed by Reid \cite{reid}.
 After which, several experiments were performed to
 demonstrate quantum steering and its asymmetry \cite{Kocsis,Saunders,Handchen,Sun}.
 Most recently, Kogias {\it et al.} \cite{Adesso2015}  proposed an operational measure of  quantum steering for bipartite Gaussian states of continuous variable systems. They found that for two-mode Gaussian states, the quantum steering reduces to a form of coherent information and the asymmetry of steering  cannot exceed $\ln 2$ \cite{Adesso2015}.

On the other hand, relativistic quantum information \cite{Peres, Schuller-Mann, RQI1,jieci1, jieci2, Ralph,RQI6,
adesso2,RQI2,RQI3,RQI4,RQI5,RQI7, adesso3}, the study of
quantum information processes and concepts  in a relativistic setting, has been a
blooming area of research  for both conceptual and experimental reasons.
Understanding quantum phenomena in a relativistic framework
is necessary because the realistic quantum systems are essentially noninertial. It was experimentally demonstrated that the gravitational frequency shift (GFS) effects have a remarkable influence on the  precision of atomic clocks for a variation of $0.33$m in height \cite{Alclock}. In addition,  relativistic effects of the Earth notably affect satellite-based quantum information processing
tasks \cite{Wangqkd,Bruschi:Ralph:14, Bruschi:Ralph:15} and quantum clock synchronization \cite{wang15}.
Quantum information also plays a prominent role in the study of the
thermodynamics and  information loss problem \cite{Bombelli-Callen,Hawking-Terashima} of black holes.
Therefore, it is of  great interest to study  how relativistic
effects influence the properties of quantum steerability \cite{Sabin2015} in a curved spacetime.

In this work we present a quantitative investigation of Gaussian quantum steerability for free bosonic modes in the background
of an eternal Schwarzschild black hole \cite{RQI4}.
We assume that Alice and Bob initially share a two-mode squeezed Gaussian state with squeezing $s$ \cite{adesso3}. Alice is a Kruskal observer who stays stationary at an asymptotically flat region (or freely falls into the black hole), while Bob is a Schwarzschild observer who hovers near the
 event horizon of the black hole  with  uniform acceleration. A vacuum state observed by
 Alice would be detected as a thermal state from Bob's viewpoint. From a general relativity viewpoint, the
 temperature $T$ of the Hawking thermal bath  depends on surface gravity $\kappa$ of the black hole.
In a quantum information scenario, such a process can be described  as a bosonic amplification channel acting on Bob's quantum state \cite{adesso2,adesso3}. 
We will calculate the
Gaussian quantum steering ${\cal G}^{A \to B}$, which  quantifies to what extent Bob's mode
can be steered by Alice's measurements, and the steering ${\cal G}^{B \to A}$, to verify the asymmetric property of steerability in the curved spacetime. 
We find that the quantum steerability between Alice and Bob
decreases with the increase of the Hawking temperature parameter $r$,
while the the steerability between Bob and anti-Bob
segregated by the event horizon is generated at the same time.
We also find that the attainment of maximal steering  asymmetry indicates the transition between one-way steerability  and both-way steerability for two-mode  Gaussian states under the influence of Hawking radiation.

The outline of the paper is as follows. In Sec. II we briefly  introduce the definition and measure of bipartite Gaussian quantum steering. In Sec. III we discuss how the Unruh-Hawking effects of the black hole can be described by a bosonic amplification channel acting on the covariance matrix of a bipartite system. In Sec. IV we study the behavior of Gaussian quantum steering and its asymmetry in background of a Schwarzschild black hole. The last section is devoted to a brief summary.
\section{Definition and measurement of Gaussian quantum steering \label{GSteering}}
Let us first briefly introduce the definition and measurement of Gaussian quantum steering. We consider a continuous variable quantum system \cite{weedbrook} represented
by $(n+m)$ bosonic modes of a bipartite system ${\rho _{AB}}$, composed of a subsystem $A$  of $n$ modes and a subsystem $B$ of $m$ modes.  For each mode $i$,  the corresponding
phase space variables can be denoted by $\hat a_i^A=\frac{\hat x_i^A+i\hat p_i^A}{\sqrt{2}}$ and $\hat a_i^B=\frac{\hat x_i^B+i\hat p_i^B}{\sqrt{2}}$.  The phase-space operators $\hat x_i^{A(B)},\,\,\hat p_i^{A(B)}$ can be grouped together into a vector
$\hat R = (\hat x_1^A,\hat p_1^A, \ldots ,\hat x_n^A,\hat p_n^A,\hat x_1^B,\hat p_1^B, \ldots ,\hat x_m^B,\hat p_m^B)^{\sf T}$, which satisfies the canonical commutation relations $[{{{\hat R}_i},{{\hat R}_j}} ] = i{\Omega _{ij}}$, with $\Omega  =  \bigoplus_1^{n+m} {{\ 0\ \ 1}\choose{-1\ 0}}$ being the symplectic form. Any Gaussian state ${\rho _{AB}}$  is completely
specified by its first and second statistical moments. The latter is a covariance matrix  with elements ${\sigma _{ij}} = \text{Tr}\big[ {{{\{ {{{\hat R}_i},{{\hat R}_j}} \}}_ + }\ {\rho _{AB}}} \big]$ and can always be
put into a block form
\begin{equation}\label{CM}
\sigma_{AB} = \left( {\begin{array}{*{20}{c}}
   A & C  \\
   {{C^{\sf T}}} & B  \\
\end{array}} \right).
\end{equation} The covariance matrix $\sigma_{AB}$ tcan describe a physical quantum state if and only if  (\textit{iff} ) it satisfies the \textit{bona fide} uncertainty principle relation
\begin{equation}\label{bonafide}
{\sigma _{AB}} + i\,({\Omega _{AB}} ) \ge 0.
\end{equation}

Now let us give the definition of  steerability.  After  Alice performs a set of measurements $\mathcal{M}_A$, the bipartite state $\rho_{AB}$ is $A\to B$ steerable (i.e., Alice can steer Bob) \textit{iff} it is \textit{not} possible
for every pair of local observables $R_A$ ( on $A$ with outcome $r_A$) and $R_B$, to express the joint probability as
$P\left( {{r_A},{r_B}|{R_A},{R_B},{\rho _{AB}}} \right) = \sum\limits_\lambda  {{\wp_\lambda }} \, \wp\left( {{r_A}|{R_A},\lambda } \right)P\left( {{r_B}|{R_B},{\rho _\lambda }} \right)$ \cite{wiseman}. In other words, at least one measurement pair ($R_A$ and $R_B$) is required to violate this expression when  ${\wp_\lambda }$ is fixed across all measurements. Here ${\wp_\lambda }$ and $\wp \left( {{r_A}|{R_A},\lambda }\right)$ are probability distributions and $P\left( {{r_B}|{R_B},{\rho _\lambda }} \right)$ is the conditional  probability distribution associated to the extra condition of being evaluated on the state $\rho_\lambda$. 
It has been shown in \cite{wiseman} that a general $(n+m)$-mode Gaussian state $\rho_{AB}$ is $A\to B$ steerable by Alice's Gaussian measurements \textit{iff} the condition
\begin{equation}\label{nonsteer}
{\sigma _{AB}} + i\,({0_A} \oplus {\Omega _B}) \ge 0,
\end{equation}
is violated. Henceforth, a violation of \eqref{nonsteer} is necessary and sufficient for the Gaussian  $A\to B$ steerability.

One can  define the Gaussian $A \to B$ steering to  quantify how much a bipartite Gaussian state $\sigma_{AB}$ is steerable by the measurements performed by Alice
\begin{equation}\label{GSAB}
{\cal G}^{A \to B}(\sigma_{AB}):=
\max\bigg\{0,\,-\sum_{j:\bar{\nu}^B_j<1} \ln(\bar{\nu}^B_j)\bigg\}\,,
\end{equation}
where $\{\bar{\nu}^B_{j}\}$ are the symplectic eigenvalues of the Schur complement of A in the covariance matrix $\sigma_{AB}$ \cite{Adesso2015}.

The $A \to B$ steering vanishes \textit{iff} the state described by $\sigma_{AB}$ is nonsteerable by Alice's measurements, and it generally quantifies the amount of by which the condition (\ref{nonsteer}) fails to be fulfilled. The Gaussian steerability  $A \to B$ acquires a particularly simple form when the steered party Bob has one mode only (i.e., $m=1$) \cite{Adesso2015}
\begin{eqnarray}
\nonumber {\cal G}^{A \to B}(\sigma_{AB}) &=&
\mbox{$\max\big\{0,\, \frac12 \ln {\frac{\det A}{\det \sigma_{AB}}}\big\}$}\\ &=& \max\big\{0,\, {\cal S}(A) - {\cal S}(\sigma_{AB})\big\}\,, \label{GS1}
\end{eqnarray}
where ${\cal S}(\sigma) = \frac12 \ln( \det \sigma)$ is  the R\'enyi-$2$ entropy  \cite{renyi}.
Similarly, a corresponding measure of Gaussian $B \to A$ steerability can be obtained by swapping the roles of $A$ and $B$, resulting in an expression like Eq. (\ref{GS1}).  Unlike quantum entanglement, the quantum steering is an asymmetric property \cite{Adesso2015}: a quantum state may be steerable from Alice to Bob, but not vice versa. In a quantum information scenario, quantum steering  corresponds to the task of entanglement distribution by an untrusted party \cite{wiseman}. If Alice and Bob share a state that is steerable from Alice to Bob, then Alice is able to convince Bob (who does not trust Alice) that their shared state is entangled by performing local measurements and classical communication \cite{wiseman}.

\section{Bosonic amplification channel description of the Hawking effect \label{model}}
In this section we will show how the Unruh-Hawking radiation of the black hole can be described by a bosonic amplification channel \cite{adesso2}, which is a Gaussian channel.
 The spacetime background near a Schwarzschild black hole is described by the metric
\begin{eqnarray}\label{matric}
ds^2&=&-(1-\frac{2M}{r}) dt^2+(1-\frac{2M}{r})^{-1} dr^2\nonumber\\&&+r^2(d\theta^2
+\sin^2\theta d\varphi^2),
\end{eqnarray}
where $M$ represents the mass of the black hole. Throughout this paper we set
$G=c=\hbar=\kappa_{B}=1$.

In the background of the black hole, a massless  bosonic field $\phi$ obeys the Klein-Gordon(K-G)  equation \cite{birelli}
 \begin{eqnarray}
\label{K-G Equation}\frac{1}{\sqrt{-g}}\frac{{\partial}}{\partial
x^{\mu}} \left(\sqrt{-g}g^{\mu\nu}\frac{\partial\phi}{\partial
x^{\nu}}\right)=0.
 \end{eqnarray}
 Solving Eq. (\ref{K-G Equation}) near the event horizon, we
obtain a set of positive-frequency outgoing modes propagating in  the regions inside and outside of the event
horizon
\begin{eqnarray}\label{inside mode}
\Phi^+_{{\Omega},\text{in}}\sim \phi(r) e^{i\omega u},
\end{eqnarray}
\begin{eqnarray}\label{outside mode}
\Phi^+_{{\Omega},\text{out}}\sim \phi(r) e^{-i\omega u},
\end{eqnarray}
where $u=t-r_{*}$ and $r_{*}=r+2M\ln\frac{r-2M}{2M}$ is the tortoise coordinate in Schwarzschild spacetime.

Eqs. (\ref{inside mode}) and
(\ref{outside mode}) can be used to  expand
 the scalar field $\Phi$  as
\begin{eqnarray}\label{First expand}
\Phi&=&\int
d\Omega[\hat{a}^{out}_{\Omega}\Phi^{+}_{{\Omega},\text{out}}
+\hat{b}^{out\dag}_\Omega
\Phi^{-}_{{\Omega},\text{out}}\nonumber\\ &+&\hat{a}^{in}_{\Omega}\Phi^{+}_{{\Omega},\text{in}}
+\hat{b}^{in\dag}_{\Omega}\Phi^{-}_{{\Omega},\text{in}}],
\end{eqnarray}
where $\hat{a}^{out}_{\Omega}$ and $\hat{b}^{out\dag}_{\Omega}$
are the bosonic particle annihilation and antiboson creation operators
acting on the state in the exterior region of the black hole, and
$\hat{a}^{in}_{\Omega}$ and $\hat{b}^{in\dag}_{\Omega}$ are
the boson annihilation and antiboson creation operators acting
on the interior region states. The  Schwarzschild vacuum $|0\rangle_S$ can be defined as $\hat{a}^{out}_{\Omega}|0\rangle_S=\hat{a}^{in}_{\Omega}|0\rangle_S=0$; therefore the modes $\Phi^{\pm}_{{\Omega},\text{out}}$ and $\Phi^{\pm}_{{\Omega},\text{in}}$ are usually called  Schwarzschild modes \cite{Fabbri, RQI4, Bruschi,jieci1, Bruschi2}.

 Making an analytic continuation for Eqs. (\ref{inside mode}) and
(\ref{outside mode}), we find a complete basis for positive energy
modes, i.e., the Kruskal modes, according to the suggestion of Domour and Ruffini \cite{D-R}. The  Kruskal modes can be used to define the Hartle-Hawking vacuum, which corresponds to the Minkowski vacuum in flat spacetime.  Then we can quantize the massless scalar field in the Schwarzschild  and Kruskal modes respectively \cite{jieci1,jieci2}, and can obtain the Bogoliubov transformations \cite{Barnett, birelli} between the modes and operators in different
coordinates. However, as performed in \cite{Bruschi}, an inertial observer Alice has the freedom to create excitations in any accessible
mode $\Omega_j,\forall j$. Hence, one cannot map a single-frequency  Kruskal mode
into a set of single frequency modes in Schwarzschild coordinates \cite{Bruschi}.  To avoid this  obstacle, we employ the Unruh basis which provides an intermediate step between the Kruskal and Schwarzschild modes. The relations between the  Unruh and Schwarzschild operators take the form
\begin{align}\label{Unruhop}
\nonumber C_{\Omega,\text{\text{R}}}=&\left(\cosh r_{\Omega}\, \hat{a}_{{\Omega},\text{out}}-\sinh r_{\Omega}\, \hat{b}^\dagger_{{\Omega},\text{in}}\right),\\*
\nonumber C_{\Omega,\text{\text{L}}}=&\left(\cosh r_{\Omega}\, \hat{a}_{{\Omega},\text{in}}-\sinh r_{\Omega}\, \hat{b}^\dagger_{{\Omega},\text{out}}\right),\\*
\nonumber D^\dagger_{\Omega,\text{\text{R}}}=&\left(-\sinh r_{\Omega}\, \hat{a}_{{\Omega},\text{out}}+\cosh r_{\Omega}\, \hat{b}^\dagger_{{\Omega},\text{in}}\right),\\*
D^\dagger_{\Omega,\text{\text{L}}}=&\left(-\sinh r_{\Omega}\, \hat{a}_{{\Omega},\text{in}}+\cosh r_{\Omega}\, \hat{b}^\dagger_{{\Omega},\text{out}}\right),
\end{align}
where $\sinh r_{\Omega}=(e^{\frac{ 2\pi\Omega}{\kappa}}-1)^{-\frac{1}{2}}$ and $\kappa$ is the surface gravity of the black hole which relates the Hawking temperature $T$ of the black hole by $T=\kappa/2\pi$ .

A generic Schwarzschild Fock state $|nm,pq\rangle_{\Omega}$ describing  both particles and antiparticles can be written as
\begin{align}\label{shortnot}
|nm,pq\rangle_{\Omega}:=\frac{\hat{a}_{{\Omega},\text{out}}^{\dagger n}}{\sqrt{n!}}\frac{\hat{b}^{\dagger m}_{\Omega,\text{in}}}{\sqrt{m!}}\frac{\hat{b}^{\dagger p}_{\Omega,\text{out}}}{\sqrt{p!}}\frac{\hat{a}^{\dagger q}_{\Omega,\text{in}}}{\sqrt{q!}}|0\rangle_S,
\end{align}
where the $\pm$ sign denotes the  particle and antiparticle respectively. This allows us to write the Unruh vacuum as   \cite{Fabbri,Bruschi2}
\begin{equation}\label{vacuumba}
|0_\Omega\rangle_\text{U}=\frac{1}{\cosh r_\Omega^{2}}\sum_{n,m=0}^{\infty}\tanh r_\Omega^{n+m}|nn,mm\rangle_{\Omega},
\end{equation}
where $|0_\Omega\rangle_\text{U}$ is a shortcut notation used to underline that each Unruh mode $\Omega$ is  mapped into a Schwarzschild mode $\Omega$.

One-particle Unruh states are defined as $|1_{j}\rangle^+_{\text{U}}=c_{\Omega,\text{U}}^\dagger|0\rangle_\text{H}$, $|1_{j}\rangle^-_{\text{U}}=d_{\Omega,\text{U}}^\dagger|0\rangle_\text{H}$
where the Unruh particle and antiparticle creation operators are defined as  linear combinations of the two Unruh operators
$
c_{\Omega,\text{U}}^\dagger=q_{\text{R}}C^\dagger_{\Omega,\text{R}}
+q_{\text{L}}C^\dagger_{\Omega,\text{L}}$ and
$d_{\Omega,\text{U}}^\dagger=q_{\text{R}}
D^\dagger_{\Omega,\text{R}}+q_{\text{L}}D^\dagger_{\Omega,\text{L}},
$
where $q_\text{\text{R}},q_\text{\text{L}}$  satisfy $|q_\text{\text{R}}|^2+|q_\text{\text{L}}|^2=1$.
The operator $c_{\Omega,\text{U}}^\dagger$
indicates the creation of a pair of particles \cite{jieci1}, i.e., a boson with mode $\Omega$ in the
exterior region and an antiboson in the interior region of the
black hole. Similarly, the create operator $d_{\Omega,\text{U}}^\dagger$
means that an antiboson and a boson are created outside and
inside the event horizon, respectively. 
The particles and antiparticles can
radiate randomly toward the inside and outside regions from the event
horizon with the total  probability $ |q_R|^2 + |q_L|^2 = 1$.
In this situation, $|q_R|=1$  means that all the particles move
toward the black hole exteriors while all the antiparticles move
to the inside region \cite{jieci1}; i.e., only particles can be detected as
Hawking radiation.
If we fix $q_R=1$ and assume that Bob has a
detector sensitive only to the particle modes,  Eq.~(\ref{vacuumba}) reduces to $|0_\Omega\rangle_\text{H}=\frac{1}{\cosh r_\Omega}\sum_{n=0}^{\infty}\tanh r_\Omega^{n}|nn\rangle_{\Omega}$ and can be described by a bosonic amplification channel \cite{adesso2, Bruschi}. Then the effect of Hawking radiation corresponds to a
two-mode squeezing operator acting on the input state  $|\psi_0\rangle_{out}$ for Bob
  \begin{eqnarray}\label{stator2}
 \nonumber\rho_{out} &=& {\rm tr}_{in} \{\hat{U}_{out,in}(r_\Omega) \big[(|\psi_0\rangle\!\langle\psi_0|)_{out} \\&& \otimes \ (|0\rangle\!\langle0|)_{in}\big] \hat{U}_{out,in}^\dagger(r_\Omega)\}\,,
 \end{eqnarray}
where $\hat{U}_{out,in}(r_\Omega)=e^{r_\Omega(\hat{b}^\dagger_{{\Omega},\text{out}}\hat{b}^\dagger_{{\Omega},\text{in}}-
\hat{a}_{{\Omega},\text{out}}\hat{a}_{{\Omega},\text{in}})}$ is the
two-mode squeezing operator. Hereafter we write $r_\Omega$ as $r$ for convenience.   It is worth noting that the  squeezing transformation $\hat{U}_{out,in}(r)$ is a Gaussian operation, it will preserve the Gaussianity of the input states. A symplectic phase-space representation of the two-mode squeezing operation $\hat{U}_{out,in}(r)$  has the form

\begin{eqnarray}\label{cmtwomode}
 S_{B,\bar B}(r)= \left(\!\!\begin{array}{cccc}
\cosh r&0&\sinh r&0\\
0&\cosh r&0&-\sinh r\\
\sinh r&0&\cosh r&0\\
0&-\sinh r&0&\cosh r
\end{array}\!\!\right).
\end{eqnarray}
\section{The effect of Hawking radiation on Gaussian quantum steerability \label{tools}}

In this paper we study a massless scalar field $\phi$ for two Unruh modes $A$ and $B$  whose state, as prepared in an inertial  frame, is initialized in a pure, entangled Gaussian two-mode squeezed state with squeezing $s$ \cite{adesso3}. The initial state can be  described from an inertial perspective, via its covariance matrix
\begin{eqnarray}\label{inAR}
\sigma^{\rm (M)}_{AB}(s)= \left(\!\!\begin{array}{cccc}
\mathcal{C}_2&0&\mathcal{S}_2&0\\
0&\mathcal{C}_2&0&-\mathcal{S}_2\\
\mathcal{S}_2&0&\mathcal{C}_2&0\\
0&-\mathcal{S}_2&0&\mathcal{C}_2
\end{array}\!\!\right),
\end{eqnarray}
where $\mathcal{C}_2=\cosh (2s)$ and $\mathcal{S}_2=\cosh (2s)$, and the state is prepared in  Unruh modes $A$ and $B$.    From Eq.~(\ref{stator2}) one can see that the change from Unruh modes to Schwarzschild modes corresponds to a two-mode squeezing operation associating to the symplectic transformation $S_{B,\bar B}(r)$. Under such transformation, mode $B$ is mapped into two sets of Schwarzschild regions, respectively for the exterior region $out$ and interior region  $in$ of the black hole.  From an inertial viewpoint, the system is bipartite, but
an extra set of modes $\bar{B}$  becomes relevant from the perspective of a Schwarzschild observer. Therefore, a complete description of the system involves three modes, mode $A$ described by Alice, mode $B$ described by the Schwarzschild observer Bob, and mode $\bar{B}$  by a  hypothetical observer anti-Bob confined inside the event horizon. The covariance matrix of the  Gaussian state describing  the complete system is given by \cite{adesso3}
\begin{eqnarray}\label{in34}
\nonumber\sigma^{\rm (a)}_{AB \bar B}(s,r) &=& \big[I_A \oplus  S_{B,\bar B}(r)\big] \big[\sigma^{\rm (M)}_{AB}(s) \oplus I_{\bar B}\big]\\&& \big[I_A \oplus  S_{B,\bar B}(r)\big]\,,
\end{eqnarray}
where $S_{B,\bar B}(r)$ is the phase-space representation of the two-mode squeezing operation, and we use the fact that the covariance matrix of a vacuum state is an identity matrix.

Because the exterior region is causally disconnected from the interior
region of the black hole, Alice and Bob cannot
access  mode $\bar B$ inside the event horizon \cite{jieci2}.  Taking the
trace over mode $\bar B$  we obtain
covariance matrix $\sigma_{AB}(s,r)$ for Alice and Bob
\begin{equation}\label{CM1}
\sigma_{AB}(s,r)= \left( {\begin{array}{*{20}{c}}
   \mathcal{A}_{AB} & \mathcal{C}_{AB}  \\
   {{\mathcal{C}_{AB}^{\sf T}}} & \mathcal{B}_{AB}  \\
\end{array}} \right),
\end{equation} with  elements $\mathcal{A}_{AB}=\cosh(2s)I_2$, $\mathcal{C}_{AB}=[\cosh(r) \sinh(2s)]Z_2$, and $\mathcal{B}_{AB}=[\cosh(2s) \cosh^2(r) + \sinh^2(r)]I_2$ with $Z_2=\left(
                       \begin{array}{cc}
                         1 & 0 \\
                         0 & -1 \\
                       \end{array}
                     \right).$
Employing Eq. (\ref{GS1}), we obtain an analytic expression of the $A \to B$ Gaussian  steering
 \begin{eqnarray}
 {\cal G}^{A \to B}(\sigma_{AB}) =
\mbox{$\max\big\{0,\,  \ln {\frac{\cosh(2s)}{\cosh^2(r) + \cosh(2 s) \sinh^2(r)}}\big\}$}. \label{GSab}
\end{eqnarray}
From Eq. (\ref{GSab}) we can see that the $A \to B$ Gaussian  steering depends not only the squeezing parameter $s$, but also the Hawking temperature  parameter $r$, this means that the Hawking radiation of the black hole will affect  the $A \to B$ steerability because  $\sinh r=(e^{\frac{ \Omega}{T}}-1)^{-\frac{1}{2}}$.

It is well known that the symmetric properties of quantum steering  is a
crucial issue.  For example  it was recently found that the quantum steerability from $A$ to $B$ is  asymmetric \cite{Adesso2015} to the $B \to A$ steerability in a Gaussian setting, which has been experimentally demonstrated in \cite{Handchen} in a flat spacetime. To obtain  understanding of this issue, we  calculate  the steerability $ {\cal G}^{B \to A}$ and check if the relation ${\cal G}^{A \to B} ={\cal G}^{B \to A}$ is satisfied  in the Schwarzschild spacetime. After some calculations  the  $B \to A$ steering is found to be
$ {\cal G}^{B \to A}(\sigma_{AB})
= \max\big\{0,\,  \ln {\frac{\cosh^2(r) \cosh(2 s)+ \sinh^2(r)}{\cosh^2(r) + \cosh(2 s) \sinh^2(r)}}\big\}$.
To check the degree of steerability asymmetric in the curved spacetime,  we define the Gaussian steering asymmetry  as
\begin{eqnarray}{\cal G}^\Delta_{AB}=|{\cal G}^{B \to A}-{\cal G}^{A \to B}|. \label{GSaab}
\end{eqnarray}

\begin{figure}
\includegraphics[scale=0.7]{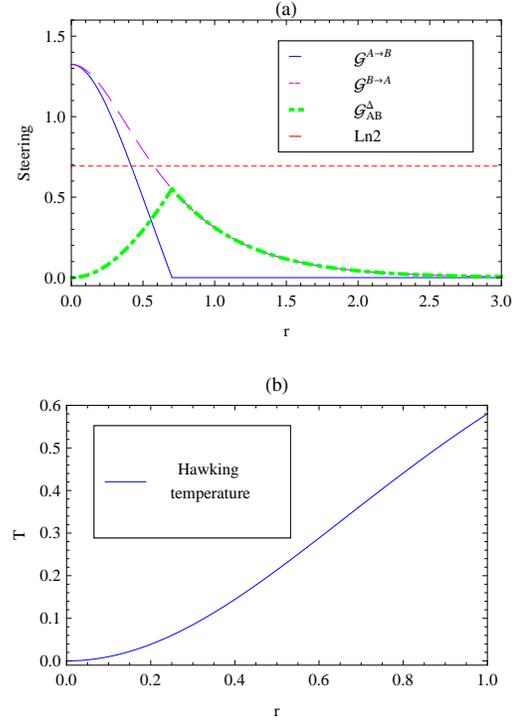}
\caption{ (a) The Gaussian quantum steering $ {\cal G}^{A \to B}$ (solid line), $ {\cal G}^{B \to A}$ (dashed line), and steering asymmetry  ${\cal G}^\Delta_{AB}$ (dot-dashed line) between Alice and Bob  as a function of the Hawking temperature  parameter $r$ of the black hole. The squeezing parameter $s$ of the initial state is fixed as $s=1$. (b) The relation between the parameter $r$ and Hawking temperature $T$ by fixing $\Omega=1$.}\label{Fig1}
\end{figure}

In Fig. (1a) we plot the steerabilities  $ {\cal G}^{B \to A}$, $ {\cal G}^{B \to A}$ as well as the Gaussian steering asymmetry  ${\cal G}^\Delta_{AB}$ as a function of the black hole's Hawking temperature  parameter $r$ for a fixed squeezing $s=1$. The relation between the parameter $r$ and Hawking temperature $T$ is given in Fig. (1b) which shows that  $T$ is a monotonically increasing function of $r$. From Fig. (1a) we can see that both  the $A \to B$ and  $B \to A$ steering decrease with  the increase of Hawking temperature parameter $r$, which means that the thermal noise introduced by Hawking effect will destroy the steerability between an inertial and an accelerated observer. It is shown that the $A \to B$ steering decreases quickly  and  suffers from a ``sudden death" with increasing  $r$, this is quite different from the behavior of quantum entanglement in a relativistic setting \cite{Schuller-Mann, adesso3,jieci2}, where  entanglement reduces to zero only at the limit of $r\rightarrow\infty$. It is shown that the $B \to A$ steering is always bigger than the $A \to B$ steering and avoids ``sudden death" with the increase of $r$, which indicates that the inertial part steers the noninertial part is easier than the noninertial part to steer the inertial part.
From Fig. (1a) we can see  that ${\cal G}^{A \to  B} \neq{\cal G}^{ B \to A}$ for any finite-valued $r$, which means  that  the  steering is always asymmetric between Alice and Bob in the curved spacetime.  The steering asymmetry increases with decreasing steerability  either way, which means that the Hawking radiation destroys the symmetry of steerability. We find that  the parameter setting  maximizing the steering asymmetry of the state $\sigma_{A B}$ is $s=arccosh(\frac{\cosh^2 r}{1-\sinh^2r})$. This condition is the same as that when the  $A \to B$ steering experiences ``sudden death" in Fig. (1a). That is, the steering asymmetry is maximal when the state is nonsteerable in the $A \to B$ way.   Therefore, the parameter $r$ attaining the peak of steering  asymmetry is that which indicates the system is currently experiencing a  transformation from both-way steerability to one-way  steerability.  In other words,  attainment of maximal steering  asymmetry indicates the transition between one-way steerability and both-way steerability for two-mode  Gaussian states under the influence of Hawking radiation. 

\begin{figure}
\includegraphics[scale=0.25]{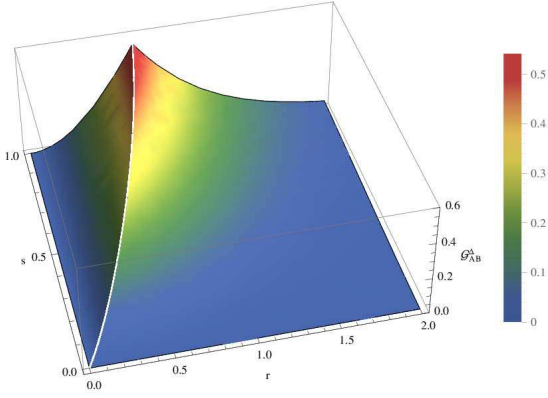}
\caption{(Color online) The Gaussian  steering asymmetry  ${\cal G}^\Delta_{AB}$ as functions of the Hawking temperature parameter $r$  and the squeezing parameter $s$.}\label{Fig2}
\end{figure}

To better understand  the interplay
between squeezing and the Hawking effect in the generation of
 Gaussian quantum steering, we plot  the  Gaussian steering asymmetry  ${\cal G}^\Delta_{AB}$ as functions of the Hawking temperature parameter $r$  and the squeezing  $s$ in Fig. (\ref{Fig2}) . It is shown  that the ${\cal G}^\Delta_{AB}$ equals  zero,  i.e., the steerability is asymmetric when $s=0$ and $r\rightarrow0$ because ${\cal G}^{A \to  B} ={\cal G}^{ B \to A}=0$ in these two cases.  The steering asymmetry monotonically  increases with increasing 
 squeezing parameter $s$, which means that the quantum resources shared in the initial state play a dominant role in quantum steering.   In addition,  the maximal steerability  point enlarges its value  increasing  $s$.

 We then study the steering between mode $B$ and mode $\bar B$ which propagate, respectively, outside and inside the event horizon.
Tracing over the modes in $A$, we obtain the covariance matrix
$\sigma_{B\bar B}(s,r)$ for Bob and anti-Bob
\begin{equation}\label{CM22}
\sigma_{B\bar B}(s,r) = \left( {\begin{array}{*{20}{c}}
   \mathcal{A}_{B\bar B} & \mathcal{C}_{B\bar B}  \\
   {{\mathcal{C}_{B\bar B}^{\sf T}}} & \mathcal{B}_{B\bar B}  \\
\end{array}} \right),
\end{equation} where  $\mathcal{A}_{B\bar B}=[\cosh(2s) \cosh^2(r) + \sinh^2(r)]I_2$, $\mathcal{C}_{B\bar B}=[\cosh^2(s) \sinh(2r)]Z_2$, and $\mathcal{B}_{B\bar B}=[\cosh^2(r) + \cosh(2s) \sinh^2(r)]I_2$.
Using  Eq. (\ref{GS1}) and  Eq. (\ref{GSaab}), we  obtain analytic expressions of the $B \to \bar B$ and $\bar B \to B$  steering, which are
 $
 {\cal G}^{B \to \bar B}(\sigma_{B\bar B}) =
\mbox{$\max\big\{0,\,  \ln [{\cosh^2(r) + \frac{\sinh^2(r)}{\cosh(2 s)}]}\big\}$}$
and
$
 {\cal G}^{\bar B \to B}(\sigma_{B\bar B}) =
\mbox{$\max\big\{0,\,  \ln[ {\sinh^2(r) + \frac{\cosh^2(r)}{\cosh(2 s)}]}\big\}$}$,
respectively. The  Gaussian  steering asymmetry  ${\cal G}^\Delta_{B\bar B}$ between Bob and anti-Bob can be computed in a similar way.

\begin{figure}
\includegraphics[scale=0.6]{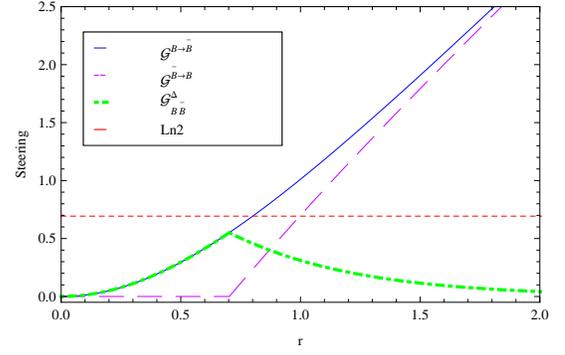}
\caption{(Color online) The Gaussian quantum steering $ {\cal G}^{B \to \bar B}$ (Solid line), $ {\cal G}^{\bar B \to B}$ (Dashed line), and steering asymmetry  ${\cal G}^\Delta_{B\bar B}$ (Dot-Dashed line) between Bob and Anti-Bob as a function of the Hawking temperature  parameter $r$ of the black hole. The squeezing parameter $s$ of the initial state is fixed as $s=1$.}\label{Fig3}
\end{figure}

In Fig. (\ref{Fig3}) we plot  the Gaussian quantum steering and  steering asymmetry between Bob and anti-Bob as a function of the Hawking temperature  parameter $r$ with fixed squeezing  $s=1$. It is shown that  quantum steerability is  generated between Bob and Anti-Bob as the increasing of the Hawking temperature parameter $r$. The steerability $ {\cal G}^{B \to \bar B}$ is nonzero for any $r$, while the steerability from anti-Bob to Bob  appears  ``sudden birth" behavior with the increase of the Hawking temperature parameter $r$.   It is interesting to find that the  maximizing  condition for the $\sigma_{B \bar B}$ steering asymmetry is $s=arccosh(\frac{\cosh^2 r}{1-\sinh^2r})$, too. Therefore, we arrive at the conclusion that  it is a  critical point of steering asymmetry in the curved spacetime. Again,  the maximal steering asymmetry  for the state $\sigma_{B \bar B}$ is obtained  when the  $\bar B \to B$ steering appears ``sudden birth".  That is to say, the parameter $r$ attaining the peak of $\bar B \to B$  steering  asymmetry is the one that indicates the system is experiencing a  transformation from one-way steerability  to both-way  steerability . Besides, we find that Bob and antiBob can steer each other  when the parameter $r$ is bigger than a critical point even though they are separated by the event horizon,  this  verifies that the  quantum steering is a nonlocal quantum correlation.
 We again find  that the quantum steering between Bob and anti-Bob is always  asymmetric for any  Hawking temperature and the maximum steering asymmetry cannot exceed $\ln2$, which  means that the state never evolves to an extremal state under the effects of Hawking radiation.

Finally, let us  present a physical
interpretation for the generation of quantum steering across the event horizon.  In this paper, anti-Bob is a hypothetical observer inside the event horizon of the black hole.  It is well known that Hawking radiation  can be explained as the spontaneous creation of  particles and
antiparticles by quantum fluctuations near the event horizon.
The particles and antiparticles will randomly
radiate  ingoing to and outgoing from the event horizon. If a particle is observed and measured by the observer Bob outside the event horizon, the state of the other antiparticle is steered and might observed by anti-Bob. Therefore, the mode $\bar B$ can be steered by measuring mode $B$ because they are pair generated and shared initial entanglement from the event horizon.

\section{Conclusions}
The effect of the Hawking effect on Gaussian quantum steering and its asymmetry in Schwarzschild spacetime are investigated. We consider three subsystems: subsystem $A$ observed by an inertial observer Alice, subsystem $B$ observed by accelerated Bob hovers near the event horizon, and subsystem $\bar B$ observed by an imaginary observer anti-Bob inside the event horizon. We obtain  a  phase-space description of quantum state evolution under the influence of the thermal bath induced by Hawking radiation.
It is shown that quantum steerability between Alice and Bob
decrease as  the Hawking temperature parameter $r$ increases. 
That is to say, thermal noise introduced by the Hawking effect will destroy the steerability between an inertial and an accelerated observer. However, the steerability between two observers segregated by the event horizon of the black hole is generated due to the effect of Hawking radiation. It is found that the steering from Alice to Bob suffers from a ``sudden death" and the steering from anti-Bob to Bob experiences as  sudden birth with the increases of Hawking temperature, which is quite different from the behavior of quantum entanglement in accelerated setting \cite{Schuller-Mann, adesso3} and curved spacetime \cite{jieci2,jieci1}. It is intriguing to find that the  steering is always asymmetric and is endowed with a maximum steering asymmetry for a fixed $r$, and that the maximum steering asymmetry  cannot exceed $\ln 2$ in the curved spacetime.  It has been shown that  $s=arccosh(\frac{\cosh^2 r}{1-\sinh^2r})$ is a  critical point of steering asymmetry under the influence of Hawking radiation. In addition, the parameters attaining the peaks of steering  asymmetry are obtained  when the $A \to B$ steering experiences ``sudden death" or  the $\bar B \to B$ steering experiences ``sudden birth".  That is to say, the attainment of maximal steering asymmetry indicates a
transition point of the two-mode Gaussian state in the Schwarzschild spacetime. These results should be significant both for giving us more information from a black hole by measuring the Hawking radiation and for our general understanding of quantum steering in a relativistic quantum system.

\begin{acknowledgments}
We acknowledge the reviewer for helpful comments, and Ioannis Kogias and Gerardo Adesso for helpful discussions. J. Wang is supported the National Natural Science Foundation
of China under Grant No. 11305058, the Doctoral Scientific Fund Project of the Ministry of Education of China under Grant No. 20134306120003, and the Postdoctoral Science Foundation of China under Grant No. 2014M560129, No. 2015T80146. J. Jing is supported the National Natural Science Foundation
of China under Grant No. 11475061. H. Fan  is supported the National Natural Science Foundation
of China under Grant No. 91536108.	
\end{acknowledgments}


\begin{thebibliography}{99}
\bibitem{schr2}
E. Schr\"odinger, Proc. Camb. Phil. Soc. {\bf 31}, 555 (1935).

\bibitem{schr}
E. Schr\"odinger, Proc. Camb. Phil. Soc. {\bf 32}, 446 (1936).

\bibitem{wiseman}
H. M. Wiseman, S. J. Jones, and A. C. Doherty, Phys. Rev. Lett.
{\bf 98}, 140402 (2007).

\bibitem{Skrzypczyk}
P. Skrzypczyk, M. Navascu\'es, and D. Cavalcanti, Phys. Rev.
Lett. {\bf 112}, 180404 (2014).

\bibitem{Walborn}
 S. P. Walborn, A. Salles, R. M. Gomes, F. Toscano, and P. H.
Souto Ribeiro, Phys. Rev. Lett. {\bf 106}, 130402 (2011).

\bibitem{Bowles}
J. Bowles, T. V\'ertesi, M. T. Quintino, and N. Brunner, Phys.
Rev. Lett. {\bf 112}, 200402 (2014).

\bibitem{steering1}
Q. Y. He, Q. H. Gong, and M. D. Reid,
Phys. Rev. Lett. {\bf 114}, 060402 (2015).

\bibitem{steering2}
C.-M. Li, K. Chen, Y.-N. Chen, Q. Zhang, Y.-A. Chen, and J.-W. Pan,
Phys. Rev. Lett. {\bf 115}, 010402 (2015).

\bibitem{steering3}
M. Marciniak, A. Rutkowski, Z. Yin, M. Horodecki, and R. Horodecki,
Phys. Rev. Lett. {\bf 115}, 170401 (2015).

\bibitem{steering4}
Q. Y. He, L. Rosales-Z\'{a}rate, G. Adesso, and M. D. Reid,
Phys. Rev. Lett. {\bf 115}, 180502 (2015).

\bibitem{steering5}
A. B. Sainz,N. Brunner, D. Cavalcanti, P. Skrzypczyk, and T. Vertesi,
Phys. Rev. Lett. {\bf 115}, 190403 (2015).

\bibitem{epr}
A. Einstein, B. Podolsky, and N. Rosen, Phys. Rev. {\bf 47}, 777
(1935).

\bibitem{eprpar}
 M. D. Reid, P. D. Drummond, W. P. Bowen, E. G. Cavalcanti,
P. K. Lam, H. A. Bachor, U. L. Andersen, and G. Leuchs, Rev.
Mod. Phys. {\bf 81}, 1727 (2009).

\bibitem{reid}
 M. D. Reid, Phys. Rev. A {\bf 40}, 913 (1989).

\bibitem{Saunders}
 D. J. Saunders, S. J. Jones, H. M. Wiseman, and G. J. Pryde,
Nat. Phys. {\bf 6}, 845 (2010).

\bibitem{Handchen}
V. Handchen, T. Eberle, S. Steinlechner, A. Samblowski,
T. Franz, R. F. Werner, and R. Schnabel, Nat. Photon. {\bf 6}, 598
(2012); S. Wollmann \emph{et al.},
Phys. Rev. Lett. {\bf116}, 160403 (2016); K. Sun \emph{et al.},
Phys. Rev. Lett. {\bf116}, 160404 (2016).

\bibitem{Sun}
 K. Sun, J.-S. Xu, X.-J. Ye, Y.-C. Wu, J.-L. Chen, C.-F. Li, and
G.-C. Guo, Phys. Rev. Lett. {\bf 113}, 140402 (2014).

\bibitem{Kocsis}
 S. Kocsis, M. J. W. Hall, A. J. Bennet, and G. J. Pryde,
Nat. Commun. {\bf 6}, 5886 (2015).



\bibitem{Adesso2015}
 I. Kogias, A. R. Lee, S. Ragy and G. Adesso, Phys. Rev. Lett. {\bf 114}, 060403 (2015).

\bibitem{Peres}
A. Peres and D. R. Terno, Rev. Mod. Phys. {\bf 76}, 93 (2004).

\bibitem{Schuller-Mann}
I. Fuentes-Schuller, and R. B. Mann, Phys. Rev. Lett. {\bf 95},
120404 (2005).

\bibitem{adesso3}
G. Adesso, I. Fuentes-Schuller, and M. Ericsson, Phys. Rev. A {\bf 76}, 062112 (2007).

\bibitem{adesso2}
M. Aspachs, G. Adesso, and I. Fuentes, Phys. Rev. Lett {\bf 105}, 151301 (2010).

\bibitem{jieci2}
J. Wang, Q. Pan, and J. Jing, Phys. Lett. B {\bf 602}, 202 (2010).

\bibitem{RQI2}
D. J. Hosler, C. van de Bruck and P. Kok, Phys. Rev. A {\bf 85}, 042312
(2012).

\bibitem{RQI5}
N. Friis, A. R. Lee, K. Truong, C. Sab\'{i}n, E. Solano, G. Johansson, and I. Fuentes, Phys. Rev. Lett. {\bf 110}, 113602 (2013).

\bibitem{RQI4}
J. Doukas, E. G. Brown, A. Dragan, and R. B. Mann, Phys. Rev. A {\bf 87}, 012306 (2013).

\bibitem{Ralph}
D. Su, and T. C. Ralph, Phys. Rev. D {\bf 90},
084022 (2014).

\bibitem{jieci1}
J. Wang, J.  Jing, and H. Fan, Phys. Rev. D {\bf 90}, 025032 (2014).

\bibitem{RQI1}
M. Ahmadi, D. E. Bruschi, and I. Fuentes, Phys. Rev. D {\bf 89},  065028  (2014);
M. Ahmadi, A. R. H. Smith, and A. Dragan,  Phys. Rev. A \textbf{92}, 062319 (2015).

\bibitem{RQI3}
A. Ch\c{e}ci\'{n}ska, and  A. Dragan, Phys. Rev. A {\bf 92}, 012321 (2015).

\bibitem{RQI6}
A. Blasco, L. J. Garay, M. Mart\'{i}n-Benito,and E. Mart\'{i}n-Mart\'{i}nez,
Phys. Rev. Lett.  {\bf 114},  141103 (2015).

\bibitem{RQI7}
B. Richter and Y. Omar, Phys. Rev. A {\bf92}, 022334 (2015).

\bibitem{Alclock}
C. W. Chou, D. B. Hume, T. Rosenband,
and D. J. Wineland,  Science \textbf{329}, 1630(2010).


 \bibitem{Wangqkd}
J. Y. Wang {\it et al.}, Nat. Photonics \textbf{10}, 387(2013).

\bibitem{Bruschi:Ralph:14}
D. E. Bruschi, T. C. Ralph, I. Fuentes, T. Jennewein, and
M. Razavi, Phys. Rev. D \textbf{90}, 045041 (2014).

\bibitem{Bruschi:Ralph:15}
D. E. Bruschi, A. Datta, R. Ursin, T. C. Ralph, and I. Fuentes,
Phys. Rev. D \textbf{90}, 124001 (2014).

\bibitem{wang15}
J. Wang, Z. Tian, J. Jing and H. Fan,  Phys. Rev. D {\bf 93}, 065008 (2016).

\bibitem{Sabin2015}
C. Sab\'in, and G. Adesso, Phys. Rev. A {\bf 92}, 042107 (2015);
D. Mondal, and C. Mukhopadhyay, arXiv:1510.07556.

\bibitem{Bombelli-Callen}
L. Bombelli, R. K. Koul, J. Lee, and R. D. Sorkin, Phys. Rev. D {\bf
34}, 373 (1986).

\bibitem{Hawking-Terashima}
S. W. Hawking, Commun. Math. Phys. {\bf 43}, 199 (1975); Phys. Rev.
D {\bf 14}, 2460 (1976); H. Terashima, Phys. Rev. D {\bf 61}, 104016
(2000).

\bibitem{weedbrook}
C. Weedbrook, S. Pirandola, R. Garc\'ia-Patr\'on, N. J. Cerf, T. C.
Ralph, J. H. Shapiro, and S. Lloyd, Rev. Mod. Phys. {\bf 84}, 621
(2012).

\bibitem{renyi}
 G. Adesso, D. Girolami, and A. Serafini, Phys. Rev. Lett. {\bf 109},
190502 (2012).

\bibitem{birelli} N. D. Birrell and P. C. W. Davies,
{\it Quantum fields in Curved Space} (Cambridge University Press,
Cambridge, 1982).


\bibitem{Fabbri}
A. Fabbri and J. Navarro-Salas, {\it Modeling Black Hole
Evaporation} (Imperial College Press, London, 2005).


\bibitem{Bruschi}
D. E. Bruschi, J. Louko, E. Mart\'{i}n-Mart\'{i}nez, A. Dragan, and I. Fuentes, Phys. Rev. A {\bf 82}, 042332 (2010).

\bibitem{Bruschi2}
D. E. Bruschi, A. Dragan, I. Fuentes, and J. Louko, Phys. Rev. D {\bf 86}, 025026 (2012).

\bibitem{D-R}
T. Damoar and R. Ruffini, Phys. Rev. D {\bf 14}, 332 (1976).

\bibitem{Barnett}
S. M. Barnett and P. M. Radmore, \it{Methods in Theoretical
Quantum Optics} (Oxford University Press, New York, 1997), pp. 67-80.

\end{thebibliography}
\end{document}